\documentstyle[epsfig]{mn}

\newif\ifAMStwofonts
    \def\lsim{\, \lower2truept\hbox{${< \atop\hbox{\raise4truept\hbox{$\sim$}}}$}\,}
\def\gsim{\, \lower2truept\hbox{${> \atop\hbox{\raise4truept\hbox{$\sim$}}}$}\,}

\title{Sunyaev-Zeldovich Effect from Quasar-driven Blast-waves}
\author[P.Platania et al.]
       {P. Platania$^1$, C. Burigana$^2$,
       G. De Zotti$^3$, E. Lazzaro$^4$, and M. Bersanelli$^1$\\
       $^1$Universit\`{a} di Milano, Dipartimento di Fisica, Via Celoria
        16, I-20133, Milano, ITALY \\
    $^2$Istituto TESRE, CNR, Via Gobetti 101, I-40129 Bologna, ITALY \\
    $^3$Osservatorio Astronomico di Padova, Vicolo dell'Osservatorio 5,
    I-35122 Padova, ITALY \\
       $^4$Istituto di Fisica del Plasma, CNR, Via Cozzi 53, I-20125,
    Milano, ITALY
    }
\date{Accepted \ldots 2002.
      Received \ldots 2002 ;
      in original form 2002 \ldots}

\pagerange{\pageref{firstpage}--\pageref{lastpage}}
\begin{document}

\maketitle

\label{firstpage}

\begin{abstract}
Quasar-driven winds are currently the best candidates for
accounting for the pre-heating of the intergalactic medium in
clusters. Such winds, occurring during early phases of the
evolution of spheroidal galaxies, shock-heat the interstellar gas,
thus inducing a detectable Sunyaev-Zeldovich effect. We estimate
the amplitude and the angular scale of such effect as well as its
counts as a function of the comptonization parameter $y$. The
contamination due to radio emission by the quasar itself is also
discussed. The corresponding mean Compton distortion of the cosmic
microwave background spectrum is found to be well below the
COBE/FIRAS upper limit.
\end{abstract}


%
% - da qualche parte di' che si calcola l'sz solo a frequenze minori di 217 GHz
% magari dove spieghi l'osservabilita', dicendo che per freq maggiori l'effetto
% si somma all' emissione del quasar e quindi non la puoi distinguere, la vedi
% come una sorgente.
% - Fig. 1, equation (5)
% - stima dell'emissione galassia ospite quando il qso e' morto

\section{Introduction}
\label{intro}

It is natural to expect that extremely powerful sources, such as
quasars, strongly affect the surrounding medium. %For example,
%quasars have long been proposed as the prime candidates for
%re-ionizing the intergalactic medium (Rees \& Setti 1970).
The recent evidences of a tight relationship between black hole
masses and velocity dispersions of the host galactic bulges
(Ferrarese \& Merritt 2000; Gebhardt et al. 2000a,b; Merritt \&
Ferrarese 2001a,b; McLure \& Dunlop 2001) have strongly
strengthened the case for a close connection between the
evolutionary pathways of spheroidal galaxies and of quasars.
Super-massive black holes have been found to be ubiquitous at the
centers of local spheroidal galaxies (Kormendy \& Richstone 1995;
Magorrian et al. 1998; van der Marel 1999). The observed
correlation between the mass of the black hole and that of the
galaxy spheroidal component hints at a substantial feedback of the
nuclear activity to the surrounding medium (Silk \& Rees 1998;
Monaco et al. 2000; Granato et al. 2001): energetic quasar-driven
winds can sweep up the interstellar gas and halt both star
formation and the growth of the central black hole.

If a significant fraction of the huge amount of energy released by
quasars %(up to $10^{62}\,$erg, and perhaps even more),
goes into ionization and heating of the neighboring gas, Compton
cooling may produce a detectable Sunyaev-Zeldovich (1972; SZ)
effect. The idea of strong shock heating of the medium originated
by energetic outflows from early quasars, originally developed by
Ikeuchi (1981), was recently revived by Natarajan, Sigurdsson \&
Silk (1998), Natarajan \& Sigurdsson (1999), and Aghanim, Balland
\& Silk (2000). Evidences of strong quasar-driven winds can be
seen in Broad Absorption Lines (BAL) quasars, comprising
10\%--15\% of optically selected quasars (Hamann \& Ferland 1999
and reference therein). The dynamic interaction of such energetic
outflows with the surrounding proto-galactic gas can heat it to
high temperature. A release of an amount of mechanical energy from
Active Galactic Nuclei several times larger than that produced by
supernovae in the surrounding galaxies may be necessary to account
for the pre-heating of the intergalactic medium in clusters (Wu et
al. 2000; Kravtsov \& Yepes 2000; Balogh et al. 2001; Bower et al.
2001). Natarajan \& Sigurdsson (1999) suggested that the SZ effect
associated to very energetic, quasar driven, winds may account for
the reported isolated Cosmic Microwave Background (CMB)
temperature decrements in directions where no clusters of galaxies
are detected, but quasar pairs are present (Jones et al. 1997;
Richards et al. 1997). Detection of this effect would obviously be
informative on the physics of quasar/galaxy evolution.

In this paper we present a new investigation of the problem,
taking explicitly into account the relevant energetics as well as
the observed relationships between black hole and host galaxy
properties. Our approach, focussing on energetics, is less liable
to uncertainties ensuing from the poor knowledge of the details of
the gas heating process. %We discuss the observability of the
%effect at microwave to mm wavelengths, in the presence of local
%radio emission and work out estimates of its number counts. % and
%of their dependence on the angular resolution of the detectors.
In Section 2 we introduce the basic ingredients entering our
calculations and estimate the amplitude and the angular scale of
the SZ effect. In Section 3 we present tentative estimates of its
number counts. The results are discussed in Section 4.

\section{SZ effect from quasars}
\label{amplitude}

The main factors determining the amplitude of the quasar-driven SZ
effect, usually measured in terms of the comptonization parameter
\begin{equation}
y = \int_{-R}^R {kT_e(r)\over m_e c^2} \sigma_T n_e(r) dr \ ,
\label{y1}
\end{equation}
(where $T_e$, $n_e$ and $m_e$ are the electron temperature,
density and mass, $\sigma_T$ is the Thomson cross section, and $R$
is the radius of the ionized region) can be illustrated as
follows.

The fractional amount of energy, $\Delta \epsilon/\epsilon_{\rm
CMB}$, transferred to the CMB by Compton cooling of the plasma is
related to $y$ by (Zeldovich \& Sunyaev 1969):
\begin{equation}
y \simeq {1\over 4} {\Delta \epsilon \over \epsilon_{\rm CMB}}\ ,
\label{y2}
\end{equation}
where $\epsilon_{\rm CMB} = a T_{\rm CMB}^4 \simeq 4.2 \times
10^{-13}(1+z)^4\,\hbox{erg}\,\hbox{cm}^{-3}$.

In the redshift range where quasars are observed ($z\lsim 6.3$,
Fan et al. 2001b) and for realistic values of the density of the
medium,
%($\Omega_{\rm IGM} \lsim 0.03$)
energetic cosmological blast waves
%containing more than $10^{57}\,$erg
remain adiabatic to relatively low $z$, i.e. the cooling time is
longer than the expansion timescale, $t_{\rm exp}$ (Voit 1996; see
also the discussion below). Thus, if a fraction $f_h$ of the total
energy released by the quasar, $E_{\rm tot}$, goes into heating of
the gas, the amount of energy per unit volume transferred to the
CMB through comptonization in a time $t < t_c$ is:
\begin{equation}
\Delta \epsilon \simeq {f_h E_{\rm tot} \over V} {t \over t_c} \ ,
\label{deltaen}
\end{equation}
where $V$ is the volume occupied by the hot gas and $t_c=3m_e
c/4\sigma_T a T_{\rm CMB}^4 \simeq 7.3 \times 10^{19}
(1+z)^{-4}\,$s is the Compton cooling timescale. We adopt, for
simplicity, an Einstein-de Sitter cosmology, so that the expansion
timescale is $t_{\rm exp}=a(t)/\dot{a}(t) = (1/H_0)(1+z)^{-3/2}$,
where $a(t)$ is the cosmic scale factor and $1/H_0 \simeq 6.17
\times 10^{17} h_{50}^{-1}\,$s,
$h_{50}=H_0/50\,\hbox{km}\,\hbox{s}^{-1}\,\hbox{Mpc}^{-1}$.

\subsection{Evolution of the shock within the quasar host galaxy}

Close relationships between the mass of the central black hole,
$M_{\rm bh}$, and the properties of host elliptical galaxies or of
galaxy bulges have been reported. Kormendy \& Richstone (1995),
Magorrian et al. (1998), and McLure \& Dunlop (2001) found that
$M_{\rm bh}$ is proportional to the luminosity of the spheroidal
galaxy component. A tighter correlation between $M_{\rm bh}$ and
the line-of-sight velocity dispersion, $\sigma$, was discovered by
Gebhardt et al. (2000) and Ferrarese \& Merritt (2000), although
the two groups find somewhat different slopes. We adopt the
relationship obtained by Gebhardt et al. (2000), based on a larger
sample:
\begin{equation}
M_{\rm bh} = 1.2\, 10^8\,\left( {\sigma \over 200\,{\rm
km/s}}\right)^{15/4}\ {\rm M}_\odot \ . \label{Mbh}
\end{equation}
The total amount of energy released by the quasar is $E_{\rm tot}
= \epsilon_R M_{\rm bh} c^2$, where $\epsilon_R \simeq 0.1$ is the
mass-to-energy conversion efficiency of the black hole.

We model the host galaxy as an isothermal sphere with density
profile
\begin{equation}
\rho(r) = \rho_0 r^{-2} \ , \label{rho}
\end{equation}
with
\begin{equation}
\rho_0 ={\sigma^2 \over 2\pi G}  \ , \label{rho0}
\end{equation}
and an outer cut-off radius $R_g$ defined by the condition that
the mean density within $R_g$ is $\delta \simeq 200$ times higher
than the mean density of the universe at the redshift $z_f$ when
the galaxy formed [numerical simulations (Cole \& Lacey 1996) show
that this radius approximately separates the virialized and infall
regions; see also Navarro et al. 1997]:
\begin{eqnarray}
\!\!\!\!\!\!\!\!\!\!\!&&R_{g} = {2\sigma \over H_0 \delta^{1/2} (1+z_f)^{3/2}}\!\! \nonumber \\
\!\!\!\!\!\!\!\!\!\!\!&&\simeq  130 h_{50}^{-1} \left({E_{\rm
tot}\over 10^{62}}\, {0.1\over
\epsilon_R}\right)^{4/15}\left({\delta \over
200}\right)^{-1/2}\left(3.5\over 1+z\right)^{3/2}\ \hbox{kpc}.
\label{Rg}
\end{eqnarray}
Several lines of evidence indicate that massive ellipticals were
already in place by $z\simeq 2.5$ (Renzini \& Cimatti 1999;
Ferguson et al. 2000; Daddi et al. 2000; Cohen 2001), although the
issue is still somewhat controversial. For $z_f\simeq 2.5$
Eq.~(\ref{Rg}) yields for $R_g$ values consistent with current
estimates for massive galaxies (see, e.g., Peebles 1993). We
assume that the distribution of baryons reflects that of dark
matter.

It is interesting to note, in passing, that this model, together
with the Gebhardt et al. (2000) relationship [Eq.~(\ref{Mbh})]
implies $M_{\rm bh} \propto M_{\rm DM}^{1.25}$, $M_{\rm DM}$ being
the total mass of the galaxy, dominated by the dark matter halo.
McLure \& Dunlop (2001) find a best-fit relationship between
black-hole mass and $R$-band luminosity of the host galaxy bulge
$M_{\rm bh} \propto L_{R}^{1.525}$. Adopting, as they do, the
relationship $M_{\rm bulge} \propto L^{1.31}$, determined by
J{\o}rgensen et al. (1996), we have $M_{\rm bh} \propto M_{\rm
bulge}^{1.16}$. If we adopt the Magorrian et al. (1998)
dependence, $M_{\rm bulge} \propto L^{1.18}$, which according to
Laor (2001) may be more appropriate, we get $M_{\rm bh} \propto
M_{\rm bulge}^{1.29}$. In both cases the observationally derived
relationship between $M_{\rm bh}$ and $M_{\rm bulge}$ is very
close to the one yielded by the present model if $M_{\rm
bulge}\propto M_{\rm DM}$.

The radius $R_s$ of a self-similar blast-wave carrying an energy
$E_b= f_h E_{\rm tot}$, expanding in a medium whose density varies as $r^{-2}$,
increases with time as (Ostriker \& McKee 1988):
\begin{equation}
R_{s}=\left[\frac{\xi E_{b}}{3 \rho_{0}} \right]^{1/3} t^{2/3} \ ,
\label{Rs}
\end{equation}
where $\xi \approx 1.5$.

Using Eqs.~(\ref{Rg}) and (\ref{Rs}) it is easily checked that,
for realistic values of the parameters, the time, $t_g$, required
for the shock front to reach the outer radius of the galaxy
\begin{eqnarray}
t_{g}\!\!\!\!&\simeq&\!\!\!\!8.9\times 10^{16} h_{50}^{-3/2}
\left({E_{\rm tot}\over 10^{62}}\right)^{1/6}\!\!
\left({\epsilon_R \over
0.1}\right)^{-2/3}\cdot \nonumber \\
&&\cdot\left({f_h \over 0.1}\right)^{-1/2} \left({\delta \over
200}\right)^{-3/4} (1+z)^{-9/4}\ \hbox{s} \label{tg}
\end{eqnarray}
is always shorter than the expansion timescale $t_{\rm exp}$.
Therefore the shocks propagate outside of the host galaxy, and can
heat up the general intra-cluster medium (ICM). However, due to
the much lower electron density in the ICM, compared to the mean
density within the galaxy, most of the SZ signal comes from within
$R_g$ (see also da Silva et al. 2001).

The main radiative cooling process for the redshifts of interest
here ($z\leq 6$) is free-free, whose cooling timescale is:
\begin{equation}
t_{\rm ff}={3 n_e kT_e \over w_{\rm ff}}\ , \label{t_ff}
\end{equation}
where $w_{\rm ff} = 1.4\, 10^{-27} T_e^{1/2} n_e^2
\bar{g}\,\hbox{erg}\,\,\hbox{s}^{-1}$ is the cooling rate of a
plasma with electron number density $n_e$ and temperature $T_e$.
The present mean ratio between baryon (mostly in stars) and dark
matter mass in massive spheroidal galaxies is estimated to be
$M_{b}/M_{DM} \simeq 0.03$ (McKay et al. 2001; Marinoni \& Hudson
2002). An upper limit to the cooling rate (and, correspondingly, a
lower limit to $t_{\rm ff}$) is obtained assuming that essentially
all such baryons were in the interstellar gas, so that
\begin{equation}
n_e(r) = 3\times 10^{-2} {\sigma^2 \over 2\pi G m_p r^2} \ ,
\label{rho}
\end{equation}
where $m_p$ is the proton mass, we find (setting $\bar{g}=1$) that
$t_{\rm ff} < t_g$ [Eq.~(\ref{tg})] for
\begin{eqnarray}
r< r_{\rm ff}\!\!\!\! &\simeq&\!\!\!\!35 h_{50}^{-3/4}\!\!
\left({T_e \over 10^6{\rm K}}\right)^{-1/4}\!\! \left({E_{\rm
tot}\over 10^{62}}\right)^{7/20}\!\!\! \left({\epsilon_R\over 0.1}
\right)^{-3/5} \cdot \nonumber \\
 & & \cdot \left({f_h \over 0.1}\right)^{-1/4} \left({\delta \over 200}\right)^{-3/8}
 (1+z)^{-9/8}\ \hbox{kpc} , \label{r_ff}
\end{eqnarray}
i.e. for only a very small fraction of the volume heated by the
blast wave [see Eq.~(\ref{Rs})]. Therefore we neglect, in the
following, the radiative losses.

\subsection{Energetics of the shock}

The fraction of power released by the quasars which goes into
heating of the gas is highly uncertain. Natarajan \& Sigurdsson
(1999) assume this fraction to amount to about half of the
bolometric luminosity. Analyses of the X-ray properties of the
intra-cluster medium (ICM) and, in particular, of the
temperature-luminosity relation (Ponman et al. 1999; Valageas \&
Silk 1999; Cavaliere et al. 2000; Tozzi \& Norman 2001; Balogh et
al. 2001) indicate the need of a substantial ``pre-heating'' of
the gas before it collapses into the cluster. As discussed by
Cavaliere \& Menci (2001), measurements of the associated SZ
effect very effectively probe the processes responsible for
non-gravitational heating of the ICM. Bower et al. (2001) have
shown that, unless the transfer of supernova energy to the ICM is
very (perhaps unrealistically) efficient, additional heating
sources, such as mechanical energy from quasar-driven winds are
required. Adopting a baryon mass corresponding to $\Omega_b =
0.025 h^{-3/2}$ ($h=H_0/100$) and a fraction of baryons converted
into stars of $f_{\rm gal} = 0.16 h^{1/2}$, Bower et al. (2001)
estimate that an energy of $\epsilon_{SN}\times 10^{49}\,$erg per
$M_\odot$ of stars formed, with $\epsilon_{SN}\simeq 1.3\hbox{--}2
h^{-1/2}= 1.8\hbox{--}2.8 h_{50}^{-1/2}$ must go into heating of
the ICM, and that most of it is unlikely to come from supernova
explosions. Other energy sources, the most obvious being quasar
winds, must be advocated.

%In the present framework, the mass in stars in a galaxy with
%velocity dispersion $\sigma$ is $M_\star = 8.4\times 10^{10}
%(\sigma/200\hbox{km}\,\hbox{s}^{-1})^3 (50/H_0)
%(1+z)^{-3/2}\,M_\odot$ [were we have taken into account that the
%dark matter density is assumed to correspond to $\Omega=1$ and the
%mass fraction in stars is $f_\star = 0.025 h^{-3/2} \times
%0.16 h^{1/2}$].
Adopting a ratio of $f_\star = 0.03$ between the mass in stars and
the gravitational mass of spheroidal galaxies we have:
\begin{eqnarray}
f_h\!\!\!\!&\simeq&\!\!\!\! 4.7 \times 10^{-2} {f_\star \over
0.03}\,{\epsilon_{SN} \over 2.8} \left({\epsilon_R \over
0.1}\right)^{-4/5}\cdot \nonumber \\
\!\!\!\!&&\!\!\!\!\cdot\left({E_{\rm tot}\over 10^{62}}
\right)^{-1/5}\!\! h_{50}^{-3/2}\!\! \left({\delta \over
200}\right)^{-1/2}\!\! \left({1+z \over 3.5}\right)^{-3/2}\
,\label{fh}
\end{eqnarray}
indicating that the mechanical power is a minor fraction of the
bolometric luminosity.

\subsection{Amplitude and angular scale of the SZ effect}

As argued in sub-section 2.1, most of the SZ signal is expected to
come from within the galaxy, so that we will set, in
Eq.~(\ref{deltaen}), $V=V_g=(4\pi/3)R_g^3$ and $t=t_g$. On the
other hand, only a minor fraction $f_g$ of the energy carried by
the blast-wave goes into heating of the gas within the galaxy
(most of it is dissipated in the ICM). Inserting in Eq.~(\ref{y2})
the above results, using for $f_h$ the expression of
Eq.~(\ref{fh}), and taking into account that, in the
Rayleigh-Jeans region $(\Delta T/T)_{\rm RJ} = -2y$ we end up with
\begin{eqnarray}
\left|\left({\Delta T \over T}\right)_{\rm RJ}\right| \simeq
2.2\times 10^{-4} {f_g\over 0.1} \left({\epsilon_{SN} \over
2.8}\right)^{1/2}\!\!\left({\epsilon_R
\over 0.1}\right)^{-4/15}\! \cdot \nonumber \\
\!\!\!\!\!\!\!\!\cdot\left({E_{\rm tot}\over 10^{62}}
\right)^{4/15}\left({\delta\over 200}\right)^{1/2}
 h_{50}^{3/4} \left({f_\star \over 0.03}\right)^{1/2} \left({1+z \over
3.5}\right)^{3/2}.\label{yfin}
\end{eqnarray}
The angular radius is $\theta_{SZ} \simeq R_g/d_A$,
\begin{equation}
d_A(z)= {2c\over H_0} {1+z - (1+z)^{1/2}\over (1+z)^2} \label{dA}
\end{equation}
being the angular diameter distance. From Eq.~(\ref{Rg}) we have:
\begin{eqnarray}
\theta_{SZ}\!\!\!\!\!&\simeq&\!\!\!\!\!17''\left({E_{\rm tot}\over
10^{62}}\right)^{4/15}\!\!  \left({\epsilon_R \over
0.1}\right)^{-4/15}\!\!  \left({1+z \over 3.5}\right)^{-3/2}
\!{d_A(2.5)\over d_A(z)}, \label{theta}
\end{eqnarray}
where $d_A(2.5) = 1595/h_{50}\,$Mpc. The amplitude and the angular
diameter of the effect, $2\theta_{SZ}\simeq 34''$, for the
reference values of the parameters, are not far from the results
reported by Richards et al. (1997): $|\delta T/T| \sim 10^{-4}$
over an area of $30''\times 65"$, and by Jones et al. (1997):
$|\delta T/T| \sim 1.4\times 10^{-4}$ in a beam of $100''\times
175"$. Since powerful high-$z$ quasars are highly clustered (Croom
et al. 2001), the larger than expected observed angular scale
might be interpreted as due to the combination of blast waves from
neighbor quasars. It should be stressed, however, that in neither
case the explanation in terms of quasar-driven blast waves is
clearly supported by observations. A more conservative explanation
is that the observed signals come from the SZ effect in previously
unknown clusters.

\begin{figure}
%\vspace{7cm}  % amount of vertical space needed
\epsfig{file=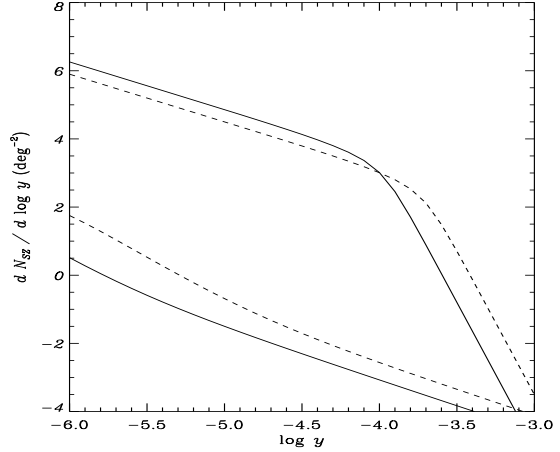,height=2.5in,width=3.1in,angle=0}
%\special{psfile=counts_bw_SZ.ps  hoffset=-50 voffset=-110
%vscale=45 hscale=45}
\caption{Differential counts per $\hbox{deg}^2$ and per unit
interval of $\log(y)$ (the two upper curves) or of $\log(y_{\rm
obs})$ (the two lower curves) of SZ signals due to quasar-driven
blast waves, as a function of the comptonization parameter $y$ (or
$y_{\rm obs}$), for the reference values of the parameters. The
two solid lines correspond to $t_q=10^7\,$yr, while the two dashed
curves correspond to $t_q=10^8\,$yr. The two lower curves take
into account the dilution effect of a Gaussian observing beam with
$\hbox{FWHM}=10'$.} \label{fig:countsSZ}
\end{figure}

\section{Counts of SZ signals from quasars}

The epoch-dependent ``luminosity function'' of SZ signals,
$\phi_{SZ}(y,z)$ can be roughly estimated from the luminosity
function of quasars, given the quasar lifetime $t_q$. We have
adopted one of the analytical evolutionary models for the B-band
luminosity function, $\phi_B(L_B,z)d\log L_B\, dz$, proposed by
Pei (1995): the two-power-law model with $h_{50}=1$, $q_0=0.5$,
and optical spectral index $\alpha_o=0.5$ ($f(\nu)\propto
\nu^{-\alpha}$). This model works well up to $z\simeq 4.5$, but
appears to under-predict by a large factor the surface density of
higher redshift quasars, which, however, are so rare (Fan et al.
2001a) that their contributions to the counts of the SZ signal is
small.

The B-band luminosity is related to the total energy released by
$E_{\rm tot} = k_B L_B t_q$, where $k_B$ is the bolometric
correction for which we adopt the value $k_B =6$ (the median value
for the sample of Elvis et al. 1994, corrected to account for the
different definition of $L_B$ used by Pei 1995). If $t_q$ is
independent of both luminosity and redshift, and considering only
the SZ effect within the host galaxy, we have:
\begin{equation}
\phi_{SZ}(y,z) = {15\over 4} \phi_B(L_B,z) {t_g\over t_q} \ ,
\label{phi_SZ}
\end{equation}
where $\phi_{SZ}(y,z)$ is the ``luminosity function" of the SZ
signal per unit $\log y$ and $z$ intervals, and the factor $15/4$
comes from $d\log L_{B}/d\log y=d\log E_{\rm tot}/d\log y$
[Eq.~(\ref{yfin})]. The quasar lifetime is still very uncertain.
Recent estimates (Salucci et al. 1999; Monaco et al. 2000; Martini
\& Weinberg 2001) suggest $t_q$ to be in the range $8\times
10^6$--$10^8\,$yr, values $\simeq \hbox{few}\times 10^7\,$yr being
favoured (Haehnelt et al. 1998; Pentericci et al. 2002; Ciotti et
al. 2001).

Inserting Eq.~(\ref{fh}) in the expression for $t_g$
[Eq.(\ref{tg})] we have:
\begin{eqnarray}
t_{g}\!\!\!\!&\simeq&\!\!\!\!7.7\times 10^{15} h_{50}^{-3/4}
\left({E_{\rm tot}\over 10^{62}}\right)^{4/15}\!\!
\left({\epsilon_R \over
0.1}\right)^{-4/15}\cdot \nonumber \\
&&\!\!\!\!\!\!\! \!\!\!\!\!\!\!\!\!\!\cdot\left({\epsilon_{SN}
\over 2.8}\right)^{-1/2}\!\! \left({f_\star \over
0.03}\right)^{-1/2}\!\!\left({\delta \over
200}\right)^{-1/2}\!\!\left({1+z \over 3.5}\right)^{-3/2} \,
\hbox{s}. \label{tg1}
\end{eqnarray}
Although the amplitude of the SZ effect is distance independent,
the observed signal is affected by distance dependent beam
dilution or resolution effects, so that the SZ signal observed
with an instrumental solid angle $\omega_{\rm beam}$ is:
\begin{equation}
y_{\rm obs}\simeq y F(\omega_{\rm beam}/\omega_{SZ}) \ ,
\label{y_obs}
\end{equation}
where $\omega_{SZ}=\pi \theta_{SZ}^2$ and
\begin{equation}
F(\omega_{\rm beam}/\omega_{SZ})\simeq \left\{ \begin{array}{ll}
\omega_{\rm beam}/\omega_{SZ} & \mbox{if $\omega_{\rm beam} <
\omega_{SZ}$} \\
\omega_{SZ}/\omega_{\rm beam} & \mbox{if $\omega_{\rm beam} >
\omega_{SZ}$}
\end{array}
\right. \ . \label{Fomega}
\end{equation}
For given $z$, $F\propto E_{\rm tot}^{\pm 8/15} \propto y^{\pm
2}$, the sign of the exponent depending on the ratio $\omega_{\rm
beam}/\omega_{SZ}$ [see Eq.~(\ref{Fomega})], so that $y_{\rm obs}
\propto y^{1\pm 2}$. The differential counts per steradian of
$y_{\rm obs}$ then write:
\begin{equation}
{dN_{SZ}\over dy_{\rm obs}} = {1\over \ln(10)\, y_{\rm obs}} \int
dz\, {dV\over dz} \phi_{SZ}(y_{\rm obs}/F,z){d\log y\over d\log
y_{\rm obs}}, \label{dN_SZ}
\end{equation}
where, for a Friedman universe, the volume element per unit solid
angle and unit $dz$ is $dV/dz=(c/H_0) d_L^2 (1+z)^{-13/2}$, $d_L=
(2c/H_0)(1+z-\sqrt{1+z})$ being the luminosity distance. The
differential counts $dN_{SZ}/d\log y = y \ln(10) dN_{SZ}/dy$ are
shown in Fig.~1 for two values of $t_q$, $10^7\,$yr and
$10^8\,$yr. For both cases, the effect of dilution by a Gaussian
beam with $\hbox{FWHM}=10'$ is also shown.

The extremely steep slope of the bright end of the counts is model
dependent. It comes out from the combined effect of the steep
slope of the high luminosity portion of the quasar luminosity
function and of the mild dependence of $y$ on $E_{\rm tot}$ [see
Eq.~(\ref{yfin})], implied by the present model. Clearly, the
assumption of dispersion-less relationships between the quasar
blue luminosity and $E_{\rm tot}$, and between $E_{\rm tot}$ and
$y$ is an over-simplification. Dispersions will result in a
flattening of the bright end of the counts.

The surface density of SZ signals is strongly dependent on the
amount of thermal energy injected in the interstellar gas of the
host galaxy, which, according to the present model, is
proportional to $t_q^{4/5}$. For $t_q$ in the range
$10^7$--$10^8\,$yr, we expect some $10^3$ SZ signals per square
degree with $y>2\times 10^{-4}$ to be detected with an imaging
survey with $\lsim 30''$ resolution (FWHM). For higher values of
$y$, the expected surface density is strongly dependent on the
quasar lifetime $t_q$. The survey should be carried out at
$10\,\hbox{GHz} < \nu < 150\,\hbox{GHz}$ to avoid blurring of the
SZ signal by local radio or dust emission. The signal is strongly
diluted in most current surveys aimed at mapping CMB anisotropies
with angular resolution $\gsim 5'$--$10'$. Even high sensitivity,
all sky surveys such as those to be carried out by ESA's {\sc
Planck} mission, cannot efficiently detect such signals. Much
better prospects for detecting SZ effects from quasar-driven blast
waves are offered by ground based bolometric arrays and
interferometers with arc-min or sub-arcmin resolution, such as the
Arcminute MicroKelving Imager (Kneissl et al. 2001), and AMiBA (Lo
et al. 2001).

Since, in the present framework, quasars are signposts of massive
galaxies at high redshifts and are therefore strongly biased
tracers of the matter distribution, we expect them (and the
associated SZ signals) to be highly clustered.

%\begin{figure}
%\vspace{5cm} \caption{Number of SZ effects from blast-waves per
%steradian as a function of $\Delta T/T$ for $t_q=10^8$ years and
%$f_k=0.5$.} \label{number}
%\end{figure}

\section{Discussion} \label{obs}

\subsection{Effect of local radio emission}

It is entirely plausible, and perhaps required by evidences of a
substantial preheating of the intra-cluster gas, that quasars
inject in the surrounding medium an amount of energy sufficient to
produce a detectable SZ effect. The corresponding angular scale is
expected to be relatively small (sub-arcmin), so that the blurring
effect by the emission from the quasar itself or from the host
galaxy may be important.

To estimate to what extent the radio emission associated with the
quasar might blur the SZ signal, we refer to the median ratio of
monochromatic luminosities at 5 GHz and at $\nu_B \simeq 6.82
\times 10^{14}\,$Hz (corresponding to $\lambda_B = 0.44\,\mu$m),
for the radio-loud and radio-quiet quasars in Table~2 of Elvis et
al. (1994). We find $\log(l_{5GHz}/l_B)_{\rm median} = -0.47$ for
radio-quiet quasars and $=2.2$ for radio loud.

Taking into account that the contribution to the antenna
temperature at the frequency $\nu$ within a solid angle
$\omega_{SZ} = \pi \theta_{SZ}^2$, of a source of 5 GHz flux
$S_{5GHz}$ and spectral index $\alpha\simeq 0.7$ ($S_\nu \propto
\nu^{-\alpha}$) is:
\begin{equation}
\Delta T_A = {S_{5GHz} (\nu/5\,{\rm GHz})^{-\alpha} c^2\over 2 k_b
\nu^2 \omega_{SZ}} \ , \label{DeltaT_A}
\end{equation}
we have, for radio-quiet quasars, in the Rayleigh-Jeans region
\begin{eqnarray}
{\Delta T_A \over T_A} &\simeq& 1.1 \, 10^{-3}\left({E_{\rm
tot}\over 10^{62}} \right)^{7/15}\left({\epsilon_R\over
0.1}\right)^{8/15}h_{50}^{2}\cdot \nonumber \\
& &\cdot {10^7 {\rm yr} \over t_q} \!\! \left({5\,{\rm GHz} \over
\nu}\right)^{2+\alpha} \left(1+z\over 3.5\right)^{-\alpha}\ ,
\label{TA}
\end{eqnarray}
In the case of radio loud quasars the coefficient would be about
500 times higher and therefore the radio emission would easily
overwhelm the SZ signal, except, perhaps, at high frequencies
($\simeq 100\,$GHz). However, only a minor fraction ($\lsim 10\%$)
of quasars are radio loud. But even in the case of radio quiet
quasars the radio emission may, at least partially, fill up the SZ
dip at $\nu \lsim 10\,$GHz, particularly if the quasar is
surrounded by an intense starburst, which is also radio bright, as
is frequently the case at high $z$ (Omont et al. 2001). The
contamination by radio emission sinks down rapidly with increasing
frequency, but at $\nu \gsim 150\,$GHz, dust emission powered by
star formation in the host galaxy, may take over, due to its
spectrum steeply rising with increasing frequency.

On the other hand, the quasar lifetime and the duration of intense
starbursts are generally shorter than the time $t_g$ for the
blast-wave to reach the boundary of the host galaxy. Therefore the
SZ signals may be observed when the quasars that originated them
are dead.

\subsection{Global comptonization distortion}

The mean distortion of the CMB spectrum due to comptonization by
the electrons heated by the blast waves, measured by the mean
value $\langle y\rangle$ of the parameter $y$ integrated along the
line of sight, is $1/4$ of the fractional amount of energy
injected in the CMB per unit volume:
\begin{equation}
\langle y\rangle = {f_{\rm h, eff}\over 4} \int \int_0^\infty dz\,
d\log L_B {\phi_B(L_B,z) k_B L_B \over \epsilon_{\rm
CMB}(z)}{t_{\rm exp}\over t_c}{dt\over dz} \ . \label{<y>}
\end{equation}
where $f_{\rm h, eff}$ is the effective fraction of the total
energy released by the quasars carried by blast waves. From the
above discussion it follows that pre-heating of the ICM requires
$f_{\rm h, eff}\lsim 0.1$. Using the double-power-law quasar
luminosity function by Pei (1995), we find:
\begin{equation}
\langle y\rangle \simeq 2.4 \times 10^{-6} {f_{\rm h, eff}\over
0.1} \ . \label{<y>max}
\end{equation}
Using Pei's (1995) exponential luminosity function, the value of
$\langle y\rangle$ increases by 20\%. On the other hand, it may be
noted that Eq.~(\ref{<y>}) includes the contribution (which turns
out to be quite substantial) of low-$z$ quasars to the integral
(the contribution from the redshift range 0--2 is comparable to
that from $2\le z \le 4$); however, in the present framework, the
most energetic blast-waves should be associated to early phases of
the quasar/galaxy evolution. Thus the expected global
comptonization distortion of the CMB spectrum induced by quasar
driven blast waves is well below the COBE/FIRAS limit $\langle
y\rangle < 1.5 \times 10^{-5}$ (Fixsen et al. 1996). This limit
could be (marginally) exceeded only in the unrealistic case that
the total energy emitted by all quasars over the entire life of
the universe went into heating of the surrounding gas. This would
also entail an excessive (by almost one order of magnitude)
pre-heating of the ICM.

\bigskip\noindent
{\bf ACKNOWLEDGMENTS}

Work supported in part by MIUR and ASI.

\bsp

\label{lastpage}

\end{document}